\documentclass[final]{siamltex}
\def \beq{\begin{equation}}
\def \eeq{\end{equation}}

\def \lf{\left(}
\def \rt{\right)}

\def \tj{{t^j\over j! }}

\title{NEW, HIGHLY ACCURATE PROPAGATOR FOR THE LINEAR AND NONLINEAR SCHR\"ODINGER EQUATION.}

\author{Hillel Tal-Ezer\thanks{School of Computer Sciences, Academic College of Tel-Aviv Yaffo,Rabenu Yeruham St., Tel-Aviv 61803, Israel, email: hillel@mta.ac.il }
                \and Ronnie Kosloff\thanks{Institute of Chemistry and
  The Fritz Haber Research Center,
  The Hebrew University, Jerusalem 91904, Israel, email: ronnie@fh.huji.ac.il }
  \and Ido Schaefer \thanks{Institute of Chemistry and
  The Fritz Haber Research Center,
  The Hebrew University, Jerusalem 91904, Israel, email: ido.schaefer@mail.huji.ac.il }}

\begin{document}

\maketitle

\begin{abstract}
A propagation method for the time dependent Schr\"odinger equation was studied leading to a general scheme
of solving ode type equations.
Standard space discretization of time-dependent pde's usually results in system of ode's of the form 
\begin{equation}
u_t - G u =s
\end{equation}
where $G$ is a operator ( matrix ) and $u$ is a time-dependent solution vector. Highly accurate methods, based on polynomial approximation of a modified exponential evolution operator, had been developed already for this type of problems
 where $G$ is a linear, time independent matrix and $s$ is a constant vector. In this paper we  will describe  a new algorithm for the more general case where $s$ is a time-dependent r.h.s vector.
 An iterative version of the new algorithm can be applied to the general case where $G$ depends on $t$ or $u$. Numerical results for  Schr\"odinger  equation with time-dependent potential and to non-linear Schr\"odinger  equation  will be presented.  
\end{abstract}



\pagestyle{myheadings}
\thispagestyle{plain}
\markboth{Hillel Tal-Ezer, Ronnie Kosloff and Ido Schaefer}{New, highly accurate propagator for the Schr\"odinger equation}

\section{Introduction}
The time dependent Schr\"odinger equation is of fundamental importance, it governs quantum dynamics. 
As a result any simulation of quantum phenomena requires an effective scheme to represent and solve this equation:
\begin{equation}\label{eq:1}
i \psi_t =  H \psi
\end{equation}
where $\psi $ is a vector representing the wave function and $H$ the Hamiltonian operator \cite{tenr}. 
Applications differ considerably. The dimension of Hilbert space required to represent the wave function $\psi$ 
can vary from 2, for a two-level-system, to $\sim 2^{30}$ in practical applications. 
When the size of Hilbert space becomes too large to be represented directly,
approximate methods are employed which lead to a nonlinear version of the Schr\"odinger equation \cite{mctd}.

The central role of Eq. (\ref{eq:1}) in quantum dynamical simulations has generated a wealth 
of numerical methods to solve the equation. For low dimensions, the common approach is based on diagonalizing the
Hamiltonian operator $H$. For higher dimensions, this becomes impractical and one has to resort to matrix-free
methods which require only the evaluation of the operation of the Hamiltonian on a vector. 
As a result, an implicit knowledge of the Hamiltonian  is sufficient.

Many methods have been developed and implemented to propagate the equation in time. 
Typically the propagation period is divided into time steps. As a result the error in each time step will accumulate.
This means that effective methods should have as large as possible time step and have a high accuracy within a time step.
For time independent Hamiltonian operators a global polynomial expansion of the propagator is the method of choice\cite{fors}:
\begin{equation}
\psi(t)=e^{-i  H t} \psi(0) \approx \sum_n a_n(t) P_n(H) \psi (0)
\end{equation} 
where $P_n(x) $ is a polynomial of order $n$ which is evaluated recursively. 
The most popular choice has been the Chebychev polynomial \cite{tal1} due to its exponential rate of convergence.
Other polynomials have been tried with similar or inferior results. 

In many applications, the Hamiltonian is explicitly time dependent. 
These include systems subject to a time dependent electromagnetic field
(spectroscopy), quantum control which requires to  infer the time dependent field 
that leads to a desired outcome such as a quantum gate\cite{jose}.
In these problems the remedy to overcome the explicit time dependence 
was to employ a short time step in which the field is approximated as  piecewise constant.
This solution immediately degrades the accuracy to first order in the time step. 
Four general approached have been explored to overcome this difficulty. 
\begin{enumerate}
\item{Solving the equations using general Taylor based solvers such as Runge Kutta or second order differencing. These methods have slow convergence properties\cite{kose}.}
\item{Employing the $(t,t')$ method which eliminates the explicit time dependence by embedding the problem in a larger Hilbert space adding  time translation
to the Hamiltonian $  H(t) \rightarrow  H(t')+i\frac{\partial}{\partial t'}$. The method restores the accuracy of the high order polynomial expansion but has been found to be expensive in use\cite{pesk}.}
\item{Another class of approaches rely on the Magnus expansion to overcome the problem of time ordering\cite{tal9}. 
The solution is cast into the form: $\psi(t) = e^{ U} \psi(0)$ and approximated as $e^{ U}\approx e^{ A_1}e^{ A_2}e^{A_3}e^{A_{...}}$. This type of solution includes the split operator method\cite{spo} as well as polynomial approximations of the exponent\cite{magn}.}
\item{ When the Hamiltonian can be split as:
$H= H_0+V(t)$ then $\psi(t) = e^{-i  H_0 t} \psi(0) -i\int_0^t e^{-i  H_0 (t-t')}   V(t')\psi(t')dt'$. This formal solution establishes the base for a polynomial approximation of the result \cite{mamd2}, \cite{mamd1}.}
\end{enumerate}

When considering nonlinear version of the Schr\"odinger equation, such as the GrossÐPitaevskii equation or time dependent density functional equations, methods 2 and 3 are not applicable and we are left with options based on 1 and 4.  The new algorithm presented in this paper belongs to the fourth approach. 
We will demonstrate that the new algorithm is highly efficient with respect to accuracy versus numerical effort,
both for linear time dependent problems as well as for non linear versions of the Schr\"odinger equation.

\section{The new algorithm (linear case)} 
Let us consider a general system of ode's of the form
\beq
u_t = Gu + s ,  \label{1}
\eeq
\beq
u(0)=v_0   ,     \label{2}
\eeq
where $G$ is a constant,  $ N \times N $ matrix. If $s$ is  constant then, by Duhamel principle, the solution is
\beq
u(t)=e^{Gt}v_0 + \int_0^t e^{G\lf t -\tau\rt}sd\tau.
\eeq
Formal integration results in
\beq
u(t)=e^{Gt}v_0+f_1\lf G,t \rt s ,
\eeq
where

\beq
 f_1\lf z, t \rt = \left\{
\begin{array}{cc}
 \frac{1}{z}\lf e^{zt}-1\rt&  z\ne 0\\
t &  z = 0 \ .
\end{array} \right.
\eeq

Since
\beq
e^{zt}=zf_1\lf z,t \rt +1 ,
\eeq
then 
\beq
e^{Gt}=Gf_1\lf G,t\rt +I
\eeq
and therefore
\beq
u(t)=v_0 + f_1(G,t)v_1
\eeq
where $v_1=Gv_0+s$. 

Going one step further, let us consider the system
\beq
u_t = Gu + s_0+ts_1.   \label{702}
\eeq
Using similar steps as above, we get the formal solution 
\beq
u(t)=v_0 + tv_1+f_2(G,t)v_2 ,
\eeq
where

\beq
 f_2\lf z, t \rt = \left\{
\begin{array}{cc}
\frac{1}{z^2}\lf e^{zt}-1-zt\rt &  z\ne 0\\
\frac{t^2}{2} &  z = 0
\end{array} \right.
\eeq

and $ v_2=Gv_1+s_1$. The following Lemma applies to the general case.

\vspace{.1in}
\noindent
\underline{\bf Lemma:}

The formal solution of the set of ode's
\beq
u_t = Gu +  \sum_{j=0}^{m-1} \tj s_j  \label{3}
\eeq

is
\beq
u = \sum_{j=0}^{m-1} \tj v_j + f_m\lf G,t\rt v_m , \label{4}
\eeq
where $ v_j $ satisfy the recurrence relation
\beq
v_0 = u_0 \label{5}
\eeq
\beq
v_j = Gv_{j-1}+ s_{j-1} \qquad \qquad 1\leq j \leq m \label{66}
\eeq
and

\beq
 f_m\lf z, t \rt = \left\{
\begin{array}{cc}
\frac{1}{z^m}\lf e^{zt}-\sum_{j=0}^{m-1}\frac{(zt)^j}{j!}\rt &  z\ne 0\\   \label{300}
\frac{t^m}{m!} &  z = 0.
\label{55}
\end{array} \right.
\eeq

\noindent
\underline{\bf Proof: }
\nopagebreak

It is easily verified that
\beq
\frac{df_m}{dt}=zf_m + \frac{t^{m-1}}{(m-1)!} \ . \label{1001}
\eeq
Hence
\beq
 u_t = \sum_{j=0}^{m-2} \tj  v_{j+1} +Gf_mv_m + \frac{t^{m-1}}{(m-1)! }v_m
\eeq
or
\beq
u_t =  \sum_{j=0}^{m-1} \tj  v_{j+1} +Gf_mv_m  \ .\label{9}
\eeq
Using ~(\ref{66}) we get
\beq
u_t = G {\sum_{j=0}^{m-1}\frac{t^j}{j!}}v_j + {\sum_{j=0}^{m-1}\frac{t^j}{j!}} s_j +Gf_m v_m  \ . \label{10}  
\eeq
Hence
\beq
u_t = G\lf {\sum_{j=0}^{m-1}\frac{t^j}{j!}}v_j +f_m v_m\rt+ {\sum_{j=0}^{m-1}\frac{t^j}{j!}} s_j   \  \label{111}  
\eeq
or
\beq
u_t=Gu+{\sum_{j=0}^{m-1}\frac{t^j}{j!}} s_j \label{13}
\eeq
and the proof is concluded.

\noindent
\underline{\bf remark:} When $z$ is very small, computing $f_m\lf z, t \rt $ as defined in  ~(\ref{55}) can be unstable  due to roundoff errors. Possible remedy is to use instead an approximation based on Taylor expansion 
\beq
f_m\lf z,t\rt = t^m \sum_{j=0}^\infty \frac{\lf zt \rt ^j}{\lf m+j \rt ! }.
\eeq
The solution vector $u$ can be approximated with high accuracy as
\beq
u\approx \sum_{j=0}^{m-1} \tj v_j + p_k\lf G,t\rt v_m \label{705}
\eeq
where $p_k(z,t)$ is 'optimal' polynomial which approximates $f_m(z,t)$ where $z\in D$ and $D$ is a domain in the complex plane which includes all the eigenvalues of $G$. The $p_k$ polynomial can be based on Chebyshev expansion ~\cite{tal1} , Arnoldi approach  ~\cite{high}, ~\cite{tal2} or Newton interpolation approach~\cite{tal3} .

In the more general case where $s$ is any function of $t$ we do first Chebyshev approximation of $s$
\beq
s\lf t \rt\approx \sum_{j=0}^{m-1}\tilde s_jT_j(t)
\eeq
and then transform the expansion to the Taylor-like representation as in  ~(\ref{3}) \cite{mamd1}.

\section{Time-Dependant $G$}

Let us consider now the case where the matrix $G$ depends on $t$

\beq
u_t = G(t)u + s(t),\qquad u^0=u(0)   ,\qquad  0 \le t \le T.  \label{11}
\eeq
(The time dependent Schr\"odinger equation where the potential depends on $t$ is an example of such an equation).

In order to apply the new algorithm in this case, one has to resort to a time-steps algorithm.
Consider  that we have marched already to time level $t_n$ and we want to compute the solution at time level $t_{n+1}$. ~(\ref{11}) can
be written as
\beq
u_t = G_nu + s_n(t)   ,\qquad  0 \le t \le T  
\eeq
where
\beq
G_n=G\lf t_n+\frac{\Delta T}{2}\rt,\qquad s_n\lf t \rt =s(t)+\lf G(t) - G_n \rt u,\label{101}
\eeq
and
\beq
 \Delta t = t_{n+1}-t_n.
\eeq
Observe that $s_n(t)$ depends on $u$ which is unknown yet at the time interval $[t_n, t_n+\Delta t ]$ but, as described in  Main Algorithm  below, a set of approximated vectors $u_n^j,\ j=1,\dots, m $ which approximate the solution at the Chebyshev time points 
\beq
t_j =t_n+\frac{\Delta t}{2}\lf 1-y_j \rt,\ \ y_j=\cos \lf \frac{ \lf j-1\rt \pi}{m-1} \rt,\qquad 1\le i \le m, 
\eeq
 
can be computed in the previous time step and is used to compute the $s_j$ vectors as defined in  ~(\ref{3}).  Only in the first step one has to use an iterative algorithm in order to compute the set of approximated solution vectors at the time points
 \beq
t_j =\frac{\Delta t}{2}\lf 1-y_j \rt,\ \ y_j=\cos \lf \frac{ \lf j-1\rt \pi}{m-1} \rt,\qquad 1\le j \le m
\eeq
where the first guess is
\beq
u^1_j=u^0,\qquad 1\le j \le m.
\eeq
The iterative algorithm  is stopped when $||u_m^{k+1}-u_m^k||$ satisfies the desired accuracy.

\vspace{.1in}

\noindent
\underline{\bf First Step Algorithm}

\vspace{.1in}

Given: $u^0,\ \epsilon,\ m $  and let $t_j =\frac{\Delta t}{2}\lf 1-\cos \lf \frac{ \lf j-1\rt \pi}{m-1} \rt \rt,\ 1\le j \le m$

1.) $u_j = u^0,\qquad j=1,\dots,m$

2.) Compute $\hat s_j = s_0(t_j) \qquad$ (defined in ~(\ref{101}))

3.) Compute $s_j$ (defined in ~(\ref{3}))  by using cosine transform of  $\hat s_j$ and 

$\qquad$ then Taylor-like transform

4.) Use  ~(\ref{5})  to compute $u_j^{\textrm{new}},\ 1\le j \le m$

5.) if $||u_m^{\textrm{new}}-u_m|| \le \epsilon $ then stop

6.) $u_j=u_j^{\textrm{new}},\qquad 1\le j \le m $

7.) go to 2

\vspace{.1in}

After computing the initial solution vectors at the time points $t_j =\frac{\Delta t}{2}\lf 1-\cos \lf \frac{ \lf j-1\rt \pi}{m-1} \rt \rt,\ 1\le j \le m$
we are ready to continue with the main algorithm  which computes the solution at the time interval $[0, T]$.

\vspace{.1in}

\noindent
\underline{\bf Main Algorithm}

\vspace{.1in}

Given: $v_0,\ m ,\ \ \{u_j\}_{j=1}^m,\ T$

 $t=0, n=0$

1.) Let $ t_j^1 = t+\frac{\Delta t}{2}\lf 1-\cos \lf \frac{ \lf j-1\rt \pi}{m-1} \rt\rt,
\ t_j^2 = t+\Delta t+\frac{\Delta t}{2}\lf 1-\cos \lf \frac{ \lf j-1\rt \pi}{m-1} \rt\rt ,$ \\ $ \qquad 1\le j \le m $

1.) Compute $\hat s_j = s_n(t_j^1), \qquad $ (defined in ~(\ref{101}))

2.) Compute $s_j$ (defined in ~(\ref{3}))  by using cosine transform of  $\hat s_j$ and 

$\qquad$ then Taylor-like transform

3.) Use  ~(\ref{5})  to compute $\{u_j\}_{j=1}^m$ at $\{t_j^2\}_{j=1}^m$

4.) if $t=T$ then stop

5.) $t=t+\Delta t,\ n=n+1 $

7.) go to 1

\vspace{.1in}

Observe that $t^1_m=t^2_1=t+\Delta t$ hence, at each step, the solution vector at this point is computed twice.  The first one is the predictor and the second one is the corrector.

\section{Nonlinearity}

Let us consider now the nonlinear case.

\beq
u_t = G(u)u + s(u)   ,\qquad  0 \le t \le T.  \label{51}
\eeq
Implementation of the new algorithm in this case is almost the same as it is done in the case described in the previous section.  ~(\ref{51}) can
be written as
\beq
u_t = G_nu + s_n   ,\qquad  0 \le t \le T  \label{211}
\eeq
where
\beq
G_n=G\lf u\lf t_n+\frac{\Delta T}{2}\rt\rt,\qquad s_n =s(u)+\lf G(u) - G_n \rt u.
\eeq
The rest of the description of the algorithm is exactly the same as in the previous section.

\section{Numerical Examples} 

The numerical examples presented in this section address the case where the eigenvalues of the spatial matrix $G$ are on the imaginery axis. In this case one can use the  Chebyshev approach   \cite{tal1}. In a future paper we will treat the more general case ( e.g. boundary value problems, advection diffusion ) where the domain of eigenvalues is on the left side of the complex plane.

\vspace{.1in}
\noindent
\underline{ Example 1: Time-dependent r.h.s}

Let us consider the differential equation
\beq
u_t=u_x + s(x,t)  \qquad 0\le x \le 2\pi
\eeq
where
\beq
s(x,t)=\sin(6x)\cos(t)-2\cos(10x)\cos(2t)-6\cos(6x)\sin(t)+10\sin(10x)\sin(2t).
\eeq
The exact solution is
\beq
u(x,t)=\sin(t)\sin(6x)+\sin(2t)\cos(10x).
\eeq
Since we have periodicity in space  we can use spectral Fourier for space approximation. It results in a set of ode's
\beq
u_t=Gu+s
\eeq
where $u$ is a vector of length $n$ ( number of grid points ), $G$ is an $n\times n$ matrix which carries out the Fourier spectral differentiation  and $s$ is a vector of length $n$ which is time-dependent.

We have solved this problem in the time interval $[0\ 5] $. Since the solution is periodic with highest mode equal to 10, using $n=32$
is suffice  to compute exactly the spatial derivative. Hence, the error comes solely from time approximation.

In order to compute solution in this time interval which satisfies
\beq
||u-u_{exact}|| \le 10^{-5}
\eeq
we had to use $m=k=14$ ( these parameters are defined in ~(\ref{3}) and ~(\ref{5}) ). It means that all together we had to do $28$ matrix-vector multiplications. 

Applying standard ODE45 for this problem, we had to do $860$ matrix-vector multiplications in order to compute the solution to the desired accuracy.

\vspace{.1in}
\noindent
\underline{ Example 2: Schr\"odinger  equation with time-dependent potential}

As a second example, we consider harmonic oscillator of mass $m=1\,$
and frequency $\omega=1\,$ driven by a linearly polarized electromagnetic field with frequency $\nu=1$. 
We have to solve 
\beq
\psi_t = -iH\lf r,t\rt\psi
\eeq
where the 
time-dependent Hamiltonian is given by
\beq  \label{eq:hamoscill}
H\lf r,t\rt = -\frac{1}{2}\frac{\partial^2}{\partial r^2}+\frac{1}{2}r^2 +
                 r \sin^2\lf\frac{\pi t}{ T }\rt \cos(t) \ .
\eeq
The final time is set to $T = 15\ $. The Hamiltonian is represented on a Fourier
grid with $n =128$ grid points, and
$r_{\max} = 10\,$ $=-r_{min}$. We have used the spectral Fourier method to approximate the spatial derivatives.

Taking the initial wave function to be 
\beq
\psi(r,0)=e^{ -r^2}
\eeq
we computed the numerical solution by two methods :

 1.) RK4 (Runge-Kutta of order 4 )
 
 2.) the new algorithm.  

In all the tables below, matvecs represents the number of applications of the Hamiltonian.

The first table presents the RK4 results. For stability , the time step should be $\Delta t =0.01$, hence the minimal number of time steps needed to march to $T=15$ is $1500$. 

\noindent
\underline{Table 1-RK4}
\nopagebreak
\vspace{.2in} 

\begin{tabular}{|c|c|c|}
\hline
time-steps & matvecs & Relative L2 Error  \\
\hline
1500 &   6000 & 5.6e-04 \\
3000 & 12000 &  3.5e-05 \\
6000 &  24000 & 2.2e-06\\
\hline 
\end{tabular}
\vspace{.2in} 

Observe that dividing  the time step by $2$, the error is reduced by a factor of almost $16$ as it should be since RK4 is a scheme of order $4$. 
Hence, in order to get high accuracy, e.g. of order $10^{-10}$  , one should do around $192000$ matrix-vector multiplications.

In the next few tables we present  the results for the new algorithm. The tables differ by the $m$ and $k$ parameters where $m$ is the number of Chebyshev points in the interval $[t, t+\Delta t ]$ and $k$ is the degree of the polynomials used to approximate the function $f_m$.

\noindent
\underline{Table 2- new algorithm, m=k=7}
\nopagebreak
\vspace{.2in} 

\begin{tabular}{|c|c|c|}
\hline
time-steps & matvecs & Relative L2 Error  \\
\hline
350 &   4563 & 3.7e-02 \\
400 & 5213 &  3.9e-08 \\
600 &  7813 & 3.9e-10\\
\hline 
\end{tabular}
\vspace{.2in}

\noindent
\underline{Table 3- new algorithm, m=k=8}
\nopagebreak
\vspace{.2in} 

\begin{tabular}{|c|c|c|}
\hline
time-steps & matvecs & Relative L2 Error  \\
\hline
300 &   4515 & 2.3e-08 \\
400 & 6015 &  1.3e-09 \\
450 &  6765 & 4.8e-10\\
\hline 
\end{tabular}
\vspace{.2in} 

\noindent
\underline{Table 4- new algorithm, m=k=9}
\nopagebreak
\vspace{.2in} 

\begin{tabular}{|c|c|c|}
\hline
time-steps & matvecs & Relative L2 Error  \\
\hline
280 &   4777 & 6.0e-08 \\
350 &   5967 & 1.4e-09 \\
400 & 6817 &  3.2e-10 \\
\hline 
\end{tabular}
\vspace{.2in} 

Remark: The minimal number of time-steps presented in the last $3$ tables were such that taking smaller number will result in instability.

Observe that the new algorithm is significantly more efficient then RK4, especially when one is interested in high accuracy. In this case, the new algorithm is almost $30$ times more efficient then RK4.

\vspace{.1in}
\noindent
\underline{ Example 3: Nonlinear Schr\"odinger  equation }

For a nonlinear example we choose  the GrossÐPitaevskii equation describing the dynamics of a
Bose-Einstein-Condensate (BEC) in a harmonic trap:
\beq
\psi_t = -iH\lf r,\psi\rt\psi
\eeq
where the  Hamiltonian is given by

\beq
  H(r,\psi) = -\frac{1}{2}\frac{\partial^2}{\partial r^2}+\frac{1}{2}r^2 + |\psi|^2.
\eeq

The final time is set to $T = 10$. The Hamiltonian is represented on a Fourier
grid with $n=128$ grid points, and
$r_{\max} = 8\sqrt { \pi } = -r_{\min}$. The spectral Fourier method is used to approximate the spatial derivatives.

The initial state is 
\beq
\psi_0=e^{i8r}v_0
\eeq
where $v_0$ is the eigenvector related to the smallest eigenvalue of the nonlinear Hamiltonian.

As in the previous example, we computed the numerical solution by RK4 and by the new algorithm.

The next table presents the RK4 results. For stability , the time step should be $\Delta t =0.01515$, hence the minimal time steps needed to march to $T=10$ is $660$. 

\noindent
\underline{Table 5-RK4}
\nopagebreak
\vspace{.2in} 

\begin{tabular}{|c|c|c|}
\hline
time-steps & matvecs & Relative L2 Error  \\
\hline
660 &   2640 & 4.96e-01 \\
1320 & 5280 &  2.00e-02 \\
2640 &  10560 & 1.20e-03\\
5280 &  21120 & 7.29e-05\\
\hline 
\end{tabular}
\vspace{.2in} 

Taking into account that RK4 is a scheme of order $4$ we can conclude that in order to get high accuracy, e.g. of order $10^{-10}$  , one should do around $382000$ matrix-vector multiplications.

In the next few tables we present  the results for the new algorithm. As in the previous example, the tables differ by the $m$ and $k$ parameters.

\noindent
\underline{Table 6 - new algorithm, m=k=7}
\nopagebreak
\vspace{.2in} 

\begin{tabular}{|c|c|c|}
\hline
time-steps & matvecs & Relative L2 Error  \\
\hline
300 &   4043 & 3.3e-05 \\
500 & 6643 &  9.5e-08 \\
700 &  9243 & 1.7e-09\\
\hline 
\end{tabular}
\vspace{.2in}

\noindent
\underline{Table 7- new algorithm, m=k=9}
\nopagebreak
\vspace{.2in} 

\begin{tabular}{|c|c|c|}
\hline
time-steps & matvecs & Relative L2 Error  \\
\hline
200 &   3587 & 5.67e-05 \\
300 &   5287 & 1.14e-07 \\
400 & 6987 &  3.00e-09 \\
500 &  8687 & 6.43e-10\\
\hline 
\end{tabular}
\vspace{.2in} 

Observe that for moderate accuracy of order $ 10^{-5},\ 21120$ matvecs were needed in the RK4 case while using the new algorithm with $m=k=9$, only $3587$ matvecs were needed. The increase in efficiency is more pronounced when high accuracy is needed. For order of $10^{-10}$ accuracy, $382000$ matvecs are needed in the RK4 case compared to $8687$ matvecs for the new algorithm.

\section{Conclusions}

In this paper we have presented a new algorithm for solving a class of linear and nonlinear Schr\"odinger equations
which can be applied to general system of ode's. In the stationary linear case it is possible to reach the upper time level in one step with very high accuracy. Due to the fact that there is only one step, the accuracy is not deteriorating since there is no accumulation of errors. In the case where the matrix involved depends on time or in the case of nonlinearity, the time interval should be divided to time steps but the size of the time step is significantly larger than what is needed in standard explicit algorithms like Runge-Kutta.

The high accuracy (spectral) of the algorithm can be traced to the fact that the algorithm does not use any Taylor considerations. Taylor theorem is an extremely important tool in analysis but due to its locality it can lead to inferior numerical approximation. We believe that whenever it is possible to develop an algorithm which is Taylor free, one should explore this possibility.

\section{Remarks}

During the refereeing process we came to know of methods known as Exponential Integrators (e.g. \cite{cali}, \cite{hoch1},\cite{hoch2},\cite{lubi} )
which also make use of the functions defined in ~\ref{300}. The algorithms described in those paper are, like Runge-Kutta approach, based on Taylor considerations while the algorithm described here is Taylor-free. This is the main difference between the two approaches. Since the present algorithm is Taylor-free, there is no meaning to the term - "order of the method"  which we have in the Exponential Integrator methods.

\end{document}